\documentclass[twoside,twocolumn,english,aps,prx]{revtex4-1}

	\usepackage[usenames,dvipsnames]{xcolor}
	\usepackage{amsmath}
	\usepackage{amsfonts}
	\usepackage{amssymb}
	\usepackage[colorlinks=true,citecolor=blue,linkcolor=red]{hyperref}
	\usepackage{graphicx}
	\usepackage{bbold}	% for \mathbb{1} as a nice unit matrix symbol
	\usepackage[makeroom]{cancel}		% crossed terms in tables
	\usepackage{multirow}				% merge cells vertically in tables
	\usepackage[normalem]{ulem}        % strike-through text
	\usepackage{array}
	\usepackage{comment}
	\usepackage{pdfpages}
	\makeatletter
	\AtBeginDocument{\let\LS@rot\@undefined}
	\makeatother

	\DeclareMathOperator{\tr}{tr}

\begin{document}

\title{Keldysh Nonlinear Sigma Model for a Free-Fermion Gas under Continuous Measurements}

\author{Qinghong Yang$^{1}$}\email{yqh19@mails.tsinghua.edu.cn}
\author{Yi Zuo$^{2}$}
\author{Dong E. Liu$^{1,3,4,5}$}\email{dongeliu@mail.tsinghua.edu.cn}

\affiliation{$^{1}$State Key Laboratory of Low Dimensional Quantum Physics, Department of Physics, Tsinghua University, Beijing, 100084, China}
\affiliation{$^{2}$Beijing National Laboratory for Condensed Matter Physics, and Institute of Physics, Chinese Academy of Sciences, Beijing 100190, China}
\affiliation{$^{3}$Beijing Academy of Quantum Information Sciences, Beijing 100193, China}
\affiliation{$^{4}$Frontier Science Center for Quantum Information, Beijing 100184, China}
\affiliation{$^{5}$Hefei National Laboratory, Hefei 230088, China}

\date{\today}

%%%%%%%%%%%%%%%%%%%%%%%%%%%%%%%%%%%%%%%%
\begin{abstract}
Quantum entanglement phase transitions have provided new insights to quantum many-body dynamics. Both disorders and measurements are found to induce similar entanglement transitions. Here, we provide a theoretical framework that unifies these two seemingly disparate concepts and discloses their internal connections. Specifically, we  analytically analyze a $d$-dimension free-fermion gas subject to continuous projective measurements. By mapping the Lindblad master equation to the functional Keldysh field theory, we develop an effective theory termed as the time-local Keldysh nonlinear sigma model, which enables us to analytically describe the physics of the monitored system. Our effective theory resembles to that used to describe the disordered fermionic systems. As an application of the effective theory, we study the transport property and obtain a Drude-form conductivity where the elastic scattering time is replaced by the inverse measurement strength. According to these similarities, two different concepts, measurements and disorders, are unified in the same theoretical framework. A numerical verification of our theory and predictions is also provided.\\
%\textbf{keywords}: measurement-induced entanglement transition, disordered systems, Keldysh field theory, Keldysh nonlinear sigma model.

\end{abstract}

\maketitle

%%%%%%%%%%%%%%%%%%%%
%%%%%%%%%%%%%%%%%%%%
\section{Introduction}
The entanglement entropy, as a characteristic measure of quantum correlations, has been intensively studied in many fields of physics~\cite{nielsen,laflorencie,calabrese,ryu}. Subsystem entanglement entropies follow distinct scaling laws 
for different dynamical phenomena in quantum many-body systems. By adjusting the system parameters, different scaling laws can be mutually converted. One typical example is the transition between the phase obeying the eigenstate thermalization hypothesis (ETH)~\cite{deutsch,srednicki} and the many-body localized (MBL) phase~\cite{fleishman,gornyi,basko,oganesyan,kjall,vosk}.
When quantum many-body systems obey ETH, the entanglement entropy of subsystems presents a volume-law scaling . By increasing the disorder strength, the systems will enter the MBL phase where the subsystem entanglement entropy obeys the area law instead~\cite{pal,bauer,luitz,schreiber,khemani,abanin}. An alternative way to obtain the entanglement transition has been proposed by using projective measurements~\cite{nahum,li}. Intuitively, one can imagine
that local projective measurements will collapse a highly entangled many-body state, thus enough measurements will convert the volume-law entangled state to an area-law one. This phenomenon has been studied in a wide variety of models~\cite{nahum,li,li2,skinner,gullans,jian,choi,szyniszewski,agrawal,cao,fuji,alberton,maimbourg,chen,goto,doggen}, and knowing the entanglement transition makes us relate the monitored systems with quantum error correction~\cite{choi}.

Based on observations from the entanglement transition, one may wonder if there are internal connections between these two different concepts, measurements and disorders. In addition, the comprehensive knowledge of monitored systems and identification of potential applications thereof are contingent upon the disclosure of other properties that are currently unexplored. For example, in the {\em disorder}-induced entanglement transition case, we also know the transport property of corresponding systems. In the MBL phase, degrees of freedom are indeed being localized, which is a manifestation of the area-law entanglement, and in turn, this results in a zero DC conductivity~\cite{basko,Basko07,gopalakrishnan,nandkishore,potter}. This property signifies that the system has the capability to maintain its primary information, thereby rendering it a noteworthy strategy for improving quantum memory. Since in the measurement-induced transition counterpart, the dynamics will also be hindered by continuous measurements and the subsystem entanglement entropy also has an area-law scaling, it is natural to ask whether an analogous  localization effect exists and what is the behavior of the conductivity.

In this work, we develop an effective theory to analytically study properties of a $d$-dimension free-fermion gas under continuous projective measurements, and focus on the underlying connection between projective measurements and disorders. In order to reveal the entanglement transition, previous studies~\cite{nahum,li,li2,skinner,gullans,jian,choi,szyniszewski,agrawal,cao,fuji,alberton,maimbourg,chen,goto,doggen} mostly focus on quantum trajectory dynamics conditioned on measurement outcomes~\cite{jacobs,breuer,alberton}.
In contrast to their calculations, our theoretical scheme directly captures the unconditional dynamics generated by the full Lindblad master equation~\cite{lindblad,breuer}. Note that if the quantity is a linear function of the system's state described by the density matrix, the conditional and the unconditional approachs will give the same result. Many physical observables including the conductivity are linear functions of states. We then modify the Keldysh field theory mapping~\cite{sieberer} to capture the Lindblad master equation for open fermionic systems. Very surprisingly, the Keldysh Lindblad partition function for the monitored case resembles to the partition function in the disordered fermionic case ~\cite{kamenev,kamenev2,horbach,liao}, although measurements and disorders look quite different in the master equation formalism (see Fig.~\ref{fig:comparison} for the comparison). Inspired by this observation, we develop an effective theory termed as the {\em time-local Keldysh nonlinear sigma model} (KNSM), to describe the physics of the monitored free-fermion gas. As an application of our effective theory, we study the transport property and obtain a Drude-form conductivity where the inverse measurement strength plays the role of the elastic scattering time. This result shows a slow-down effect or diffusive behavior~\cite{znidaric,znidaric2,turkeshi,jin,jin2} due to measurements.

Sec. \ref{sec:M} sets the model under our consideration. The Keldysh Lindblad partition function of the model is given in Sec. \ref{sec:KLPF}. The effective KNSM and the time-local diffuson are obtained in Sec. \ref{sec:KNSM} and Sec. \ref{sec:TD}, respectively. In Sec. \ref{sec:DCC}, DC conductivity is derived from KNSM, and in Sec. \ref{sec:NV}, we perform a numerical verification for our theory. Appendices \ref{sec:LMEtKFT}--\ref{sec:BEFaKNSM} give detailed derivations of some formulas in the main text.

%%%%%%%%%%%%%%%%%%%%
%%%%%%%%%%%%%%%%%%%%
\section{model}\label{sec:M}

We consider a $d$-dimension spinless free-fermion gas, whose Hamiltonian reads 
\begin{equation}
    H=\int d\mathbf{x}\,c^{\dagger}\left(\mathbf{x}\right)\left(-\frac{1}{2m}\nabla^{2}-\epsilon_F\right)c\left(\mathbf{x}\right),
\end{equation}
where $c$ $(c^{\dagger})$ is the annihilation (creation) operator of fermions, $m$ is the mass of fermions, and $\epsilon_F$ is the Fermi energy which equals to the chemical potential. This free-fermion gas is subject to continuous projective measurements, in which the projective operations can be represented by the fermion density operator $n(\mathbf{x})=c^{\dagger}(\mathbf{x})c(\mathbf{x})$. Note that $n(\mathbf{x})$ satisfies $n(\mathbf{x})(a|0_{\mathbf{x}}\rangle+b|1_{\mathbf{x}}\rangle)\propto|1_{\mathbf{x}}\rangle$ and $n^2(\mathbf{x})=n(\mathbf{x})$. For a unconditional continuous measurement process, it can be described by the Lindblad master equation \cite{fuji,jacobs}. Thus, for our case, the quantum jump operator in the Lindblad master equation is the density operator $n(\mathbf{x})$, and we have
\begin{equation}\label{eq:LME}
   \partial_{t}\rho=-i\left[H,\rho\right]+\gamma\int d\mathbf{x}\left[n(\mathbf{x})\,\rho\, n(\mathbf{x})-\frac{1}{2}\left\{ n(\mathbf{x}),\rho\right\} \right],
\end{equation}
where $\rho$ is the density matrix of the free-fermion system, and $\gamma$ is the measurement strength, which has the energy dimension and is assumed to be uniform over the space. Intuitively, the measurement strength $\gamma$ can be regarded as the number of measurement events in a unit time interval. For convenience of following treatments, the initial state is chosen to be the thermal state $\rho_0=\exp[-\beta\sum_{
\mathbf{k}}c_{\mathbf{k}}^{\dagger}(\epsilon_{\mathbf{k}}-\epsilon_F)c_{\mathbf{k}}]$ with $\beta$ being the inverse temperature. Note that the Lindblad master equation for the unconditional measurement process also describes the effect of dephasing noise, thus our following results also have insights for open quantum systems.  \\

%%%%%%%%%%%%%%%%%%%%
%%%%%%%%%%%%%%%%%%%%
\section{Keldysh Lindblad partition function}\label{sec:KLPF} 
In order to do analytical analyses, instead of focusing on the master equation formalism, we resort to the functional Keldysh field theory~\cite{kamenev,sieberer}. Following the procedures provided in Ref.~\cite{sieberer}, one can transform the fermionic Lindblad master equation Eq.~\eqref{eq:LME} to a Keldysh Lindblad partition function, which reads (see App. \ref{sec:LMEtKFT} for more details and differences compared with Ref.~\cite{sieberer})
\begin{equation}\label{eq:KMPF}
\begin{split}
  Z=\int\!\mathcal{D}\left[\psi\right]\exp&\left\{iS_{0} -\frac{\gamma}{2}\int \!dx\pmb{[}\bar{\psi}_{a}\left(x\right)\psi_{a}\left(x\right)\bar{\psi}_{b}\left(x\right)\psi_{b}\left(x\right)\right.\\
  &\qquad \qquad\qquad\;\left.\vphantom{\frac{\gamma}{2}\int dx}-\bar{\psi}_{a}\left(x\right)\hat{\tau}_{1}^{ab}\psi_{b}\left(x\right)
  \pmb{]}\right\},
\end{split}
\end{equation}
where $S_0$ is the free-fermion action in the $2\times 2$ Keldysh space, $x=(\mathbf{x},t)$ throughout the paper, $a,b\in\{1,2\}$ are the Keldysh indices, and the repeated indices imply the summation over all possible values throughout the paper. Here, $\mathcal{D}[\psi]\equiv\mathcal{D}[\bar{\psi}_{1},\psi_{1},\bar{\psi}_{2},\psi_{2}]$ with $\psi_{a}$ $(\bar{\psi}_{a})$ are Grassmann numbers after the Keldysh-Lakin-Ovchinnikov transformation~\cite{kamenev},  and $\hat{\tau}_{\mu}$ with $\mu=0,1,2,3$ are the identity and three Pauli matrices in the Keldysh space. Since $Z\equiv\tr(\rho_f)$, where $\rho_f$ is the density matrix of the final state, the normalization condition $Z=1$ is self-evident in the Keldysh formalism. In the following treatment, the time contour is chosen to be $(-\infty,+\infty)$, such that all information of the system's evolution is imprinted in the partition function. To check the normalization condition for Eq.~\eqref{eq:KMPF}, one can expand the partition function in powers of the measurement strength $\gamma$, and treat each order with the help of Wick's theorem. By doing so, one will find that in order to preserve the normalization condition, at least in the first order, the bare Green's function of free fermions should be in its full form, that is 
\begin{equation}\label{eq:FGF}
    \hat{G}(\mathbf{k};t,t^\prime)=\begin{bmatrix}G_0^{R}\left(\mathbf{k};t,t^{\prime}\right) & G_0^{K}\left(\mathbf{k};t,t^{\prime}\right)\\
0 & G_0^{A}\left(\mathbf{k};t,t^{\prime}\right)
\end{bmatrix}-\frac{i}{2}\begin{bmatrix}0 & 1\\
1 & 0
\end{bmatrix}\delta_{t,t^\prime},
\end{equation}
where $G_0^{R/A/K}$ are three typical bare Green's functions used in the standard Keldysh field theory~\cite{kamenev,altland}, and $\delta_{t,t^{\prime}}$ is the Kronecker delta symbol, which comes from the discrete time version $\delta_{j,j^{\prime}}$ with $j,j^{\prime}$ standing for the $j$th time slice and $j^{\prime}$th time slice. 
%The term $\propto \delta_{t,t^{\prime}}$ is missed in Ref.~\cite{jin}, and thus in fact their Keldysh Lindblad partition functions are unnormalized. 
Note that in the traditional Keldysh partition function derived from the Hamiltonian of a closed system~\cite{altland,kamenev}, the extra term  $\propto\delta_{t,t^\prime}$ also exists. However, one usually omits it. One argument is the $t=t^\prime$ line is a manifold of measure zero and omitting it  is inconsequential for most purposes~\cite{kamenev}. In our case, we emphasize that this $\delta_{t,t'}$ term cannot be directly omitted  due to the normalization condition mentioned above.

\begin{figure*}[t]
    \centering    \includegraphics[height=12cm]{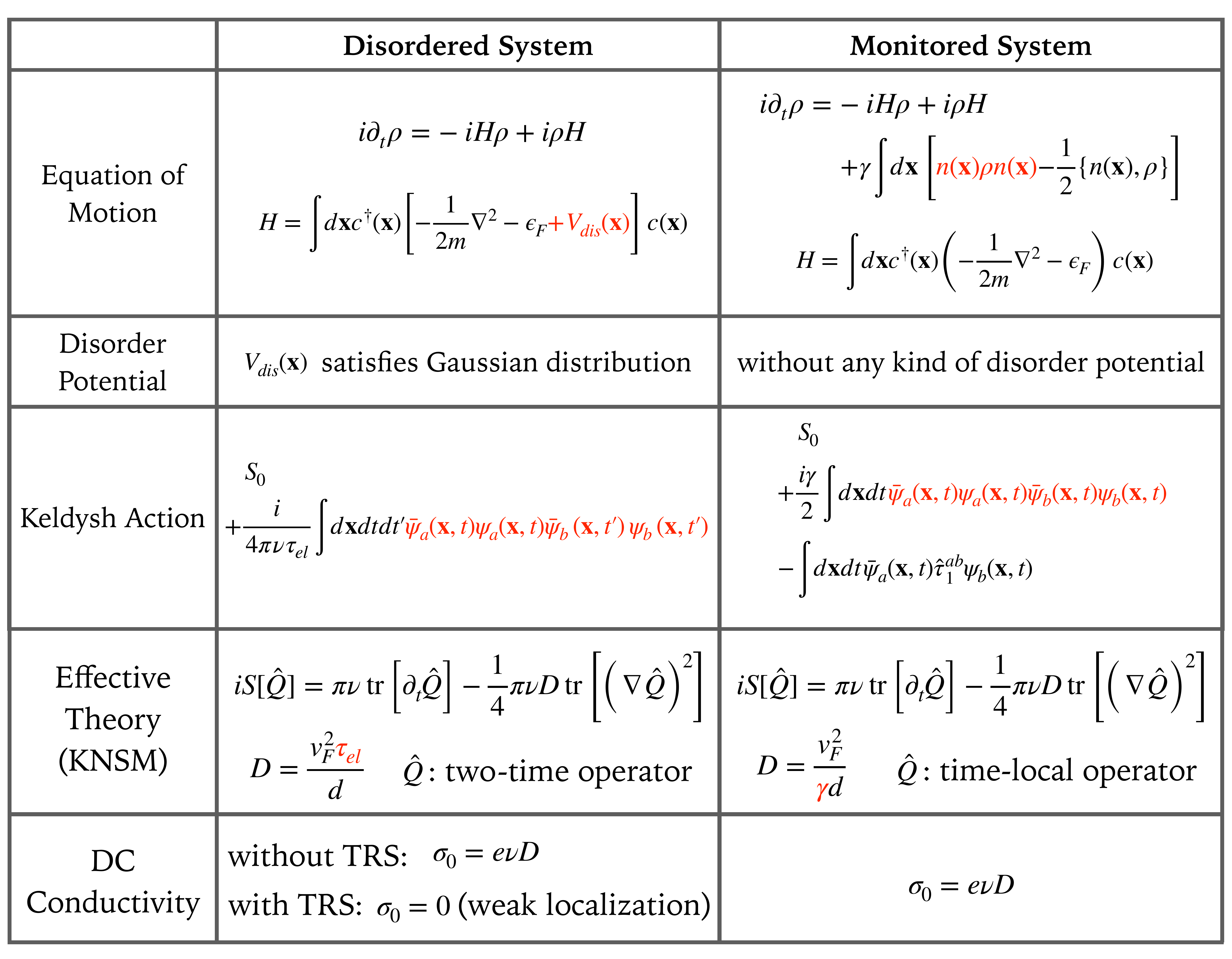}
    \caption{Comparisons between the disordered system and the monitored system in the unconditional case. One finds that the connection between measurements and disorders is not obvious in the master equation formalism, and can {\em only} be found when one resorts to the Keldysh path integral formalism. KNSM is the abbreviation of Keldysh nonlinear sigma model and TRS is that of time-reversal symmetry.}
    \label{fig:comparison}
\end{figure*}

In the Keldysh Lindblad partition function Eq.~\eqref{eq:KMPF}, we add two extra terms $\bar{\psi}_{a}\left(x\right)\psi_{a}\left(x\right)\bar{\psi}_{b}\left(x\right)\psi_{b}\left(x\right)$ with $a=b$, which are null due to the property of Grassmann numbers --- $\bar{\psi}_a^2=\psi^2_a=0$. After adding these two terms, one finds that the four-fermion term in the partition function is similar with the four-fermion term after doing disorder average in the Keldysh treatment of the disordered fermionic problem~\cite{kamenev,kamenev2,horbach,liao} (see also App. \ref{sec:KToDFS} for a brief introduction). We emphasize that such a similarity is not obvious in the master equation formalism, and can {\em only} be found when one resorts to the Keldysh path integral formalism (see Fig.~\ref{fig:comparison}).
%This inspires us to deal with this problem following the approach~\cite{horbach,kamenev,liao} used to treat the disordered fermionic system. 
However, there are also some differences between these two problems. For example, the four-fermion term in Eq.~\eqref{eq:KMPF} only depends on one time variable, while in the disordered fermionic problem, the four-fermion term depends on two time variables. In addition, there is no time-reversal symmetry in our case due to the nature of open quantum systems, while the time-reversal symmetry is present in the free-fermion gas with disorders (see App. \ref{sec:KToDFS} for detailed discussions).\\

%%%%%%%%%%%%%%%%%%%%
%%%%%%%%%%%%%%%%%%%%
\section{Time-local Keldysh nonlinear sigma model}\label{sec:KNSM}
We then try to derive an effective theory to capture and analyze the monitored system. To this end, we employ the Hubbard-Stratonovich (HS) transformation ~\cite{kamenev,altland,yqh} by introducing a {\em time-local} bosonic field $\hat{Q}$ to decouple the four-fermion term, where $\hat{Q}$ is defined as  
\begin{equation*}
 \hat{Q}=\int dx\begin{bmatrix}Q^{11}\left(x\right) & Q^{12}\left(x\right)\\
Q^{21}\left(x\right) & Q^{22}\left(x\right)
\end{bmatrix}|x\rangle\langle x|,
\end{equation*}
and it is Hermitian in the Keldysh space, i.e., $Q^{ab}(x)=[Q^{ba}(x)]^*$. Note that due to the fact that the four-fermion term depends on two time variables in the disordered fermionic case, the matrix HS field there is not diagonal in the time basis ({\em time-nonlocal}). The HS transformation and Gaussian integral lead the partition function Eq.~\eqref{eq:KMPF} to an effective bosonic theory (see App. \ref{sec:BEFaKNSM} for details):
\begin{equation}\label{eq:EBT}
\begin{split}
 Z=\int\mathcal{D}\left[\hat{Q}\right]\exp
    &\left\{ -\frac{\gamma}{2}\left(\pi\nu\right)^{2}\tr\left[\hat{Q}^{2}+\left(\frac{1}{2\pi\nu}\hat{\tau}_{1}\right)^{2}\right]\right.\\
    &\quad\left.+\tr\ln\left(-i\hat{G}^{-1}_{0}+\gamma\pi\nu\hat{Q}\right)\vphantom{\left[\hat{Q}^{2}+\left(\frac{1}{2\pi\nu}\hat{\tau}_{1}\right)^{2}\right]}\right\},
\end{split}
\end{equation}
where $\tr$ stands for the trace over the Keldysh space as well as time and spatial integrations, $\nu$ is density of states (DOS) in the vicinity of the Fermi surface and $\hat{G}_0^{-1}$ is the inverse of $\hat{G}+(i/2)\delta_{t,t^{\prime}}\hat{\tau}_1$ (see Eq.~\eqref{eq:FGF}). In the procedure of replacing $\hat{G}^{-1}$ with $\hat{G}_0^{-1}$, we have employed the argument that the $t=t^{\prime}$ line is a manifold of measure zero to higher-order ($\ge 2$) terms of $\gamma$. As mentioned in the previous, the time-reversal symmetry is absent in our case, thus we just decouple the four-fermion term in the density channel. In contrast, in the disordered fermionic case, one can also decouple the four-fermion term in the Cooper channel, and this procedure results in Cooperons, which accounts for the weak localization effect in the one-loop level of the KNSM~\cite{finkel,finkelstein,kamenev,kamenev2,horbach} (see App. \ref{sec:KToDFS} for discussions about the absence of weak localization in measurement case).

To proceed, we need to find the saddle point configuration of the action in Eq.~\eqref{eq:EBT}, which contributes most to the functional integral. Taking the variation over $\hat{Q}(x)$, one gets the saddle point equation:
\begin{equation}
    \gamma\pi^{2}\nu^{2}\hat{Q}\left(x\right)=\gamma\pi\nu\left(-i\hat{G}_{0}^{-1}+\gamma\pi\nu\hat{Q}\right)^{-1}\left(x,x\right).
\end{equation}
One can check that the constant configuration $\hat{\Lambda}=\frac{1}{2\pi\nu}\hat{\tau}_3$, satisfies the saddle point equation when $\gamma$ satisfies $\gamma\ll\epsilon_{F}$. Note that this condition also validates the procedure of replacing $\hat{G}^{-1}$ with $\hat{G}_0^{-1}$ in Eq.~\eqref{eq:EBT}. Fluctuations around the saddle point can be classified into two classes: the massive and the massless modes. For large-scale physics, the dynamics is mostly contributed by the massless modes. Thus, we here focus on fluctuations of the $\hat{Q}$-matrix along the massless ``direction", and they can be generated through the similarity transformation:
$\hat{Q}=\hat{\mathcal{R}}^{-1}\hat{\Lambda}\hat{\mathcal{R}}$. In the spacetime basis, $\hat{Q}(x)=\hat{\mathcal{R}}^{-1}(x)\,\hat{\Lambda}\,\hat{\mathcal{R}}(x)$, and $\hat{Q}(x)$ satisfies the nonlinear constraint: $\hat{Q}^2(x)=(\frac{1}{2\pi\nu})^2\hat{\tau}_0$.

In order to derive an effective theory for the massless modes, one can further employ the gradient expansion, that is, we expand the $\tr\ln$ term in Eq.~\eqref{eq:EBT} in powers of $\partial_t\hat{\mathcal{R}}^{-1}$ and $\nabla\hat{\mathcal{R}}^{-1}$. Keeping terms up to the first order of $\partial_t\hat{\mathcal{R}}^{-1}$ and the second order of $\nabla\hat{\mathcal{R}}^{-1}$, one arrives at the time-local Keldysh nonlinear sigma model (see App. \ref{sec:BEFaKNSM} for details):
\begin{equation}\label{eq:AoNLSM}
    iS\left[\hat{Q}\right]=\pi\nu\tr\left[\partial_{t}\hat{Q}\right]-\frac{1}{4}\pi\nu D\tr\left[\left(\nabla\hat{Q}\right)^{2}\right],
\end{equation}
where we just keep those non-constant terms in the action. Here, $\hat{Q}$ is redefined as $\hat{Q}=\hat{\mathcal{U}}^{-1}\hat{\mathcal{R}}^{-1}\hat{\tau}_3\hat{\mathcal{R}}\hat{\mathcal{U}}$, where $\hat{\mathcal{U}}$ encodes the statistical information and is defined as 
\begin{equation*}
    \hat{\mathcal{U}}^{-1}= \hat{\mathcal{U}}=\sum_{\epsilon}\begin{bmatrix}1 & F_{\epsilon}\\
0 & -1
\end{bmatrix}|\epsilon\rangle\langle\epsilon|
\end{equation*}
with $F_{\epsilon}=\tanh(\beta\epsilon/2)$ relating to the Fermi-Dirac distribution. The statistical distribution comes from the initial condition $\rho_0$. In Eq.~\eqref{eq:AoNLSM}, the constant $D$ is defined as  $D=v_F^2/(\gamma d)$ with $v_F$ being the Fermi velocity, and is named as the modified diffusive constant. Comparing with the traditional diffusive constant in the disordered fermionic systems, one finds that the inverse measurement strength $1/\gamma$ plays the role of the elastic scattering time (see Fig.~\ref{fig:comparison}). Intuitively, this makes sense, as the elastic scattering time represents the mean time within which a fermion hits the disorder, or in other words, is measured by the disorder. Associating with the fact that the disordered fermionic system is also described by a similar nonlinear sigma model, we know that the effect of the projective measurements has some similarities with that of disorders. Indeed, in the following, we will show that up to the one-loop level of the time-local KNSM, the conductivity is presented in the familiar Drude form~\cite{altland,kamenev}. 
Note that in Ref.~\cite{jin,jin2}, authors consider a relevant problem in one-dimension and ladder systems. They use the perturbation theory within the self-consistent Born approximation. In fact, their treatment is the saddle point of our time-local KNSM~\cite{altland}, and the similarity between projective measurements and disorders can not be seen in their treatment.\\

%%%%%%%%%%%%%%%%%%%%
%%%%%%%%%%%%%%%%%%%%
\section{Gaussian fluctuation and time-local diffuson}\label{sec:TD}
Having derived the saddle point and the effective theory for our problem, we are now in a position to draw the consequences from our effective theory. To this end, we write the similarity transformation matrix $\hat{\mathcal{R}}$ through its generator $\hat{\mathcal{W}}$ as $\hat{\mathcal{R}}=\exp(\hat{\mathcal{W}}/2)$. In the spacetime basis, we have $\hat{\mathcal{R}}(x)=\exp[\hat{\mathcal{W}}(x)/2]$. To generate a non-trivial transformation for $\hat{\tau}_3$, the generator $\hat{\mathcal{W}}(x)$ should be an off-diagonal matrix in the Keldysh space, and can be expressed as 
\begin{equation}\label{eq:GoR}
    \hat{\mathcal{W}}\left(x\right)=\begin{bmatrix}0 & d^{12}\left(x\right)\\
d^{21}\left(x\right) & 0
\end{bmatrix},
\end{equation}
where $\{\hat{\mathcal{W}}(x),\hat{\tau}_3\}=0$, and $d^{12}$ and $d^{21}$ are two independent fields. Substituting Eq.~\eqref{eq:GoR} into the nonlinear sigma model Eq.~\eqref{eq:AoNLSM}, and expanding the action in powers of $d^{12}$ and $d^{21}$, up to the second order, one obtains the Gaussian action
\begin{equation}\label{eq:GA}
 iS\left[d^{12},d^{21}\right]=\pi\nu\int dx\,d^{21}\left(x\right)\left(\partial_{t}-\frac{1}{2} D\nabla^{2}\right)d^{12}\left(x\right).
\end{equation}
With the help of the Fourier transformation, one finds that this Gaussian action will generate two types of correlators --- $\langle d^{12}_{\mathbf{k},\epsilon}d^{21}_{-\mathbf{k},-\epsilon}\rangle$ and $\langle d^{21}_{\mathbf{k},\epsilon}d^{12}_{-\mathbf{k},-\epsilon}\rangle$, which are defined as 
\begin{equation}\label{eq:MD}
\begin{split}
    \langle d^{12}_{\mathbf{k},\epsilon}d^{21}_{-\mathbf{k},-\epsilon}\rangle&=-\frac{1}{\pi\nu}\frac{1}{D^{\prime}\mathbf{k}^{2}-i\epsilon},\\\langle d^{21}_{\mathbf{k},\epsilon}d^{12}_{-\mathbf{k},-\epsilon}\rangle&=-\frac{1}{\pi\nu}\frac{1}{D^{\prime}\mathbf{k}^{2}+i\epsilon},
\end{split}    
\end{equation}
where $D^{\prime}\equiv(1/2)D$, and $\langle\cdot\rangle$ stands for taking expectation values with weight $\exp(iS[d^{12},d^{21}])$. We name these two correlators in Eq.~\eqref{eq:MD} time-local diffusons, as they are similar with those diffusons in the disorder fermionic systems~\cite{kamenev,kamenev2,horbach}. The time-local diffusons play the role of bare Green's functions and serve as the starting point to consider higher-order interaction effects and other phenomena underneath~\cite{liao,patel}.\\

%%%%%%%%%%%%%%%%%%%%
%%%%%%%%%%%%%%%%%%%%
\section{ Linear response: DC conductivity} \label{sec:DCC}
Although the evolution according to the Lindbald master equation with Hermitian jump operators will result in featureless steady state~\cite{fuji,buchhold}, due to the projection nature of the quantum jump operator $n(\mathbf{x})$, one can imagine that continuous projective measurements will have some impacts on the linear response. Here, we consider the most common linear response function in the condensed matter theory: the conductivity. For this purpose, we introduce the vector potential $\mathbf{A}(x)$, to which the current couples, through the action $S_{\mathbf{A}}=-\int dx\,\bar{\psi}_{a}(x)\mathbf{v}_{F}\mathbf{A}^{\alpha}(x)\hat{\tau}_{\alpha}^{ab}\psi_{b}(x)$~\cite{kamenev,kamenev2,horbach}, where $a,b\in\{1,2\}$, $\alpha\in\{0,1\}$,  and $\mathbf{A}^0$ stands for the classical component of the vector potential while $\mathbf{A}^1$ for the quantum component after the Keldysh transformation. Since the vector potential is classical, the quantum component $\mathbf{A}^1$ is actually zero. In the Keldysh field theory, it is preserved to generate observables by appropriate variations and is set to zero in the end. Following the procedures of deriving Eq.~\eqref{eq:AoNLSM}, one can get the KNSM in the presence of the vector potential:
\begin{equation}
    iS\left[\hat{Q},\mathbf{A}\right]=\pi\nu\tr\left[\partial_{t}\hat{Q}\right]-\frac{1}{4}\pi\nu D\tr\left[\left(\hat{\partial}\hat{Q}\right)^{2}\right],
\end{equation}
where we have assumed that the vector potential is small enough such that it does not alter the previous saddle point, $\hat{\partial}\hat{Q}=\nabla\hat{Q}+i\left[\mathbf{A}^{\alpha}\hat{\tau}_{\alpha},\hat{Q}\right]$, and $\hat{Q}$ is also defined as $\hat{Q}=\hat{\mathcal{U}}^{-1}\hat{\mathcal{R}}^{-1}\hat{\tau}_3\hat{\mathcal{R}}\hat{\mathcal{U}}$.

The longitudinal AC conductivity can be derived through $\sigma(\mathbf{q},\omega)=(-i/\omega)K^R(\mathbf{q},\omega)$, where $K^R(\mathbf{q},\omega)$ is the retarded current-current response function, and is defined as 
\begin{equation}  K^{R}\left(\mathbf{q},\omega\right)=\left.\frac{e^{2}}{2i}\frac{\delta^{2}Z\left[\mathbf{A}\right]}{\delta \mathbf{A}^{0}\left(\mathbf{q},\omega\right)\delta \mathbf{A}^{1}\left(-\mathbf{q},-\omega\right)}\right|_{\mathbf{A}=0} 
\end{equation}
with $e$ being the electron charge and $Z[\mathbf{A}]$ now being $Z[\mathbf{A}]=\int \mathcal{D}[\hat{Q}]\exp\{iS[\hat{Q},\mathbf{A}]\}$. To calculate the retarded current-current response function, one may expand $Z[\mathbf{A}]$ in powers of $\mathbf{A}$ and keep terms up to the second order of $\mathbf{A}$. Then, one finds that, up to the one-loop level of the nonlinear sigma model (Eq.~\eqref{eq:GA}), the longitudinal DC conductivity for the spatially-uniform vector potential, reads 
\begin{equation}
    \sigma\left(\mathbf{q}\rightarrow 0,\omega\rightarrow 0\right)=e^2\nu D.
\end{equation}
Thus, we reproduce the conductivity of the Drude form in a monitored free-fermion gas. Note that for a purely free-fermion gas, the conductivity is infinite, but in a monitored free-fermion gas, the conductivity is finite and is inversely proportional to the measurement strength $\gamma$. This Drude-form conductivity in the monitored free-fermion gas again presents the similarity with disorders (see Fig.~\ref{fig:comparison}). 

As mentioned previously, the absence of the weak localization effect in the measurement case results from the lack of time-reversal symmetry (TRS). According to the Drude-form conductivity, one can also obtain an intuitive picture of the lack of weak localization as following. Even though measurements and disorders will have similar effect as indicated by the similar form in the functional Keldysh field theory, the measurement case will introduce an extra decoherence effect due to the nature of open systems (and this is the origin of the lack of TRS). This decoherence effect will turn the quantum system into a classical one, and then analogous exotic effects, such as weak localization and many-body localization, are hidden in the measurement case, as even a classical system with disorders will not be localized and only Drude conductivity will be derived. Therefore, in order to reveal the localization effect produced by measurements, one has to suppress the decoherence effect by some means. This will be explored in the following work.
 
%%%%%%%%%%%%%%%%%%%%
%%%%%%%%%%%%%%%%%%%%
\section{Numerical Verification}\label{sec:NV}
To support the theory and verify our predictions, we provide a numerical test based on a one-dimensional discrete free-fermion gas subject to continuous measurements. The Hamiltonian of the one-dimensional free-fermion gas reads 
\begin{equation}
    H = \sum_{i=1}^{N-1}\;t(c^{\dagger}_{i+1}c_{i}+c^\dagger_{i}c_{i+1}),
\end{equation}
where $N$ is the number of sites, and $t$ is the hopping strength, which sets an energy scale similar with the Fermi energy in the continuum free-fermion gas model. The value of $N$ does not change the $1/\gamma$ scaling shown in the following, thus we take $N=6$ for simplicity.
The evolution is governed by 
\begin{equation}\label{eq:EoFFG}
    \partial_{t} \rho=-i[H, \rho]+\gamma \sum_i^N\left[n_i\,\rho\,n_i-\frac{1}{2}\{n_i, \rho\}\right],
\end{equation}
where $n_i=c^{\dagger}_ic_i$ is the local particle number operator. In the following, we will let $t=1$ for simplicity. Thus, the condition in our work $\gamma\ll \epsilon_F$ becomes $\gamma\ll1$ in this discrete model. In order to calculate the conductivity, we introduce the source and drain in the dissipator, in analogy to the chemical potential difference in the electrical transport experiment. Then, the master equation becomes 
\begin{equation}
\begin{split}
    \partial_{t} \rho=-i[H, \rho]&+\gamma \sum_i^N\left[n_i\,\rho\,n_i-\frac{1}{2}\{n_i, \rho\}\right]\\
    &+\gamma_s\left[c_1^{\dagger}\,\rho\,c_1-\frac{1}{2}\{c_1c^{\dagger}_1, \rho\}\right]\\
    &+\gamma_d \left[c_N\,\rho\,c^{\dagger}_N-\frac{1}{2}\{c^{\dagger}_Nc_N, \rho\}\right],\\
\end{split}
\end{equation}
where $\gamma_s$ and $\gamma_d$ are the strengths of pump and loss, respectively. The pump process simulates a source, while the loss process simulate a drain. In order to study the system described by Eq. \eqref{eq:EoFFG}, $\gamma_s$ and $\gamma_d$ should be very small, or else the property of our considered system will be changed due to those additional dissipation processes. The current operator between two neighboring sites is defined as $J_{i,i+1}=i(c_i^{\dagger}c_{i+1}-c_ic^{\dagger}_{i+1})$, and the expectation value of $J_{i,i+1}$ or the particle current $\langle J_{i,i+1}\rangle$ at time $t$ can be calculated through $\langle J_{i,i+1}\rangle(t)=\tr\left[\rho(t)J_{i,i+1}\right]$, where $\rho(t)$ is the state of the considered system at time $t$. 

According to the Fick's law~\cite{znidaric2,jin}, the particle current can also be calculated through $\langle J\rangle=-D\nabla \langle n(\mathbf{x})\rangle$, where $D$ is the diffusion coefficient and $\nabla \langle n(\mathbf{x})\rangle$ is particle number gradient. It is similar with the Ohm's law $\langle J_e\rangle=-\sigma \nabla V(\mathbf{x})$ with $\sigma$ being the electrical conductivity and $\nabla V(\mathbf{x})$ being the electrical potential gradient. Once we ignore the electron charge, the electrical potential gradient $\nabla V(\mathbf{x})$ reduces to the particle number gradient $\nabla \langle n(\mathbf{x})\rangle$, and the electrical current $\langle J_e\rangle$ becomes the particle current $\langle J\rangle$. Therefore, once we verify the $1/\gamma$ scaling of the diffusion coefficient from the Fick's law, the $1/\gamma$ scaling of the conductivity is also verified.

\begin{figure}[t]
	\centerline{\includegraphics[height=6.0cm]{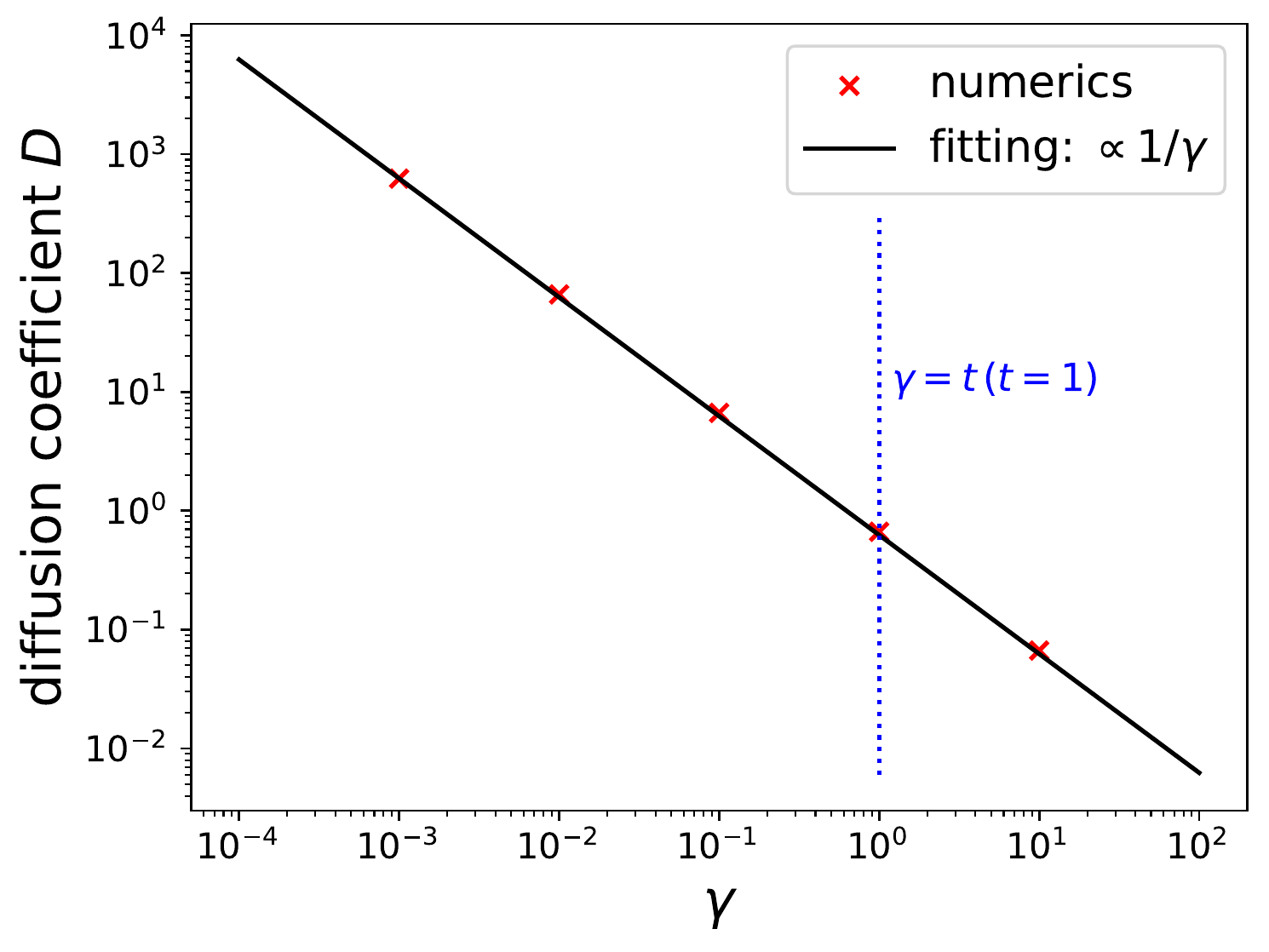}}
	\caption{Numerically calculating the diffusion coefficient $D$ of  a one-dimension discrete free-fermion gas. This figure is in the log-log plot. The red crosses are results from the numerics, and the black straight line is a fitting curve of those numerical data points. The fitting function in the log-log plot is a straight line, thus the diffusion coefficient $D$ has a $1/\gamma$ scaling behavior. In addition, according to the numerical results, the $1/\gamma$ scaling holds not only for the small $\gamma$ ($\gamma\ll t$) case, but also for the large $\gamma$ ($\gamma \gtrsim t$) case. In this numerics, we set $t=1$ for simplicity. 
		\label{fig:IGS}}
\end{figure}

We numerically solving the Lindblad master equation with source and drain, and then calculated the particle current $\langle J_{i,i+1}\rangle$ through $\langle J_{i,i+1}\rangle(t)=\tr\left[\rho(t)J_{i,i+1}\right]$. We find that the pump and loss will produce a {\em non-zero} steady particle current through the free-fermion chain for arbitrary finite measurement strength $\gamma$, and thus $\langle J_{i,i+1}\rangle$ for different $i$ are the same in the steady state. This indicates that finite measurement strength will not result in the localization effect. Without loss of generality, we choose $i=1$. Therefore, in the discrete version, after reaching the steady state, the Fick's law can be simplified as $\langle J_{1,2}\rangle=-D\left(\langle n_1\rangle-\langle n_N\rangle\right)/N$. $\langle n_1\rangle$ and $\langle n_N\rangle$ correspond to the left chemical potential and the right chemical potential, respectively, in the experiment of measuring DC conductivity. By numerically calculating $\langle J_{1,2}\rangle$, $\langle n_1\rangle$, and $\langle n_N\rangle$ for different $\gamma$, we obtain Fig.~\ref{fig:IGS}. The $\gamma\ll t$ case (in analogy with $\gamma\ll \epsilon_F$ in the continuum free fermion model), where our methods works, along with the $\gamma\gtrsim t$ case, are both considered in the numerics. We find that the diffusion coefficients of two cases both take on a perfect $1/\gamma$ scaling. This not only confirms the correctness of our theory and the used approximations, but also implies that our theory may be able to predict qualitative properties of the considered system in an extended parameter regime. The $1/\gamma$ scaling behavior indicates that for an {\em infinite} measurement strength ($\gamma\rightarrow\infty$), the system will be localized. This is a manifestation of the quantum Zeno effect~\cite{misra}.

Actually, those data points slightly deviate from the perfect $1/\gamma$ curve, but in the log-log plot, this deviation can hardly be seen. Comparing with our result, this deviation should come from higher order terms and fast varying modes.

%%%%%%%%%%%%%%%%%%%%
%%%%%%%%%%%%%%%%%%%%
\section{Discussion and conclusion}

In summary, we have derived a time-local KNSM for a free-fermion gas under continuous projective measurements. Up to the one-loop level of the effective theory, we obtain a Drude-form conductivity which is inversely proportional to the measurement strength $\gamma$, and this shows that the projective measurements cause a slow-down effect on the free-fermion gas. Interestingly, the projective measurements manifest in a form that is comparable to that of the disorders in the framework of the Keldysh field theory.
Nevertheless, the original Lindblad master equation formalism does not explicitly show this connection. Thus, in some sense, these two different concept, measurements and disorders, are unified in the framework of KNSM. Note that in the disordered fermionic system case, the weak localization effect exists in the one-loop level due to the time-reversal symmetry~\cite{kamenev,horbach}, while in our case, we do not see the weak localization effect in the one-loop level. The comparisons are summaried in Fig.~\ref{fig:comparison}. Numerical tests further confirm our theory and predictions.
%{\color{blue} Numerical tests not only support our theory and predictions, but also imply that our theory may be able to predict some qualitative properties of the monitored system with parameters out of the application range.} 
For thoroughly understanding the monitored system,  other transport properties, hydrodynamics, and quantum chaoticity are also need to be considered. Our theory is a promising method to analytically study them, and we leave these to further works.
\\

%\textbf{\large{Code availability}}

%The source codes for the numerical calculation can be found in the supplementary section $4$~\cite{supp}.\\
%\textbf{\large{Conflict of interest}}

%The authors declare that they have no conflict of interest.\\

\begin{acknowledgments}

Authors thank Jing-Yuan Chen and Sebastian Diehl for helpful discussions on the functional Keldysh field theory obtaining from Lindblad master equations, and Jing-Yuan Chen and Yunxiang Liao for helpful discussions on the Keldysh nonlinear sigma model in disordered fermionic systems. The work is supported by supported by the Innovation Program for Quantum Science and Technology (Grant No.~2021ZD0302400),
the National Natural Science Foundation of China (Grant No. 11974198).\\
\end{acknowledgments}

%\textbf{\large{Author contributions}}

%Qinghong Yang and Yi Zuo carried out the calculations; Qinghong Yang, Yi Zuo, and Dong E Liu prepared the manuscript; Qinghong Yang and Dong E Liu initialized and  supervised the project.\\

%%%%%%%%%%%%%%%%%%%%
%%%%%%%%%%%%%%%%%%%%
\begin{appendix}
\section{From Lindblad Master Equation to Keldysh Field Theory}\label{sec:LMEtKFT}

In order to introduce the mapping between the Lindblad master equation and the Keldysh field theory, we consider a trivial one-site case. The detailed procedure can be found in Ref.~\cite{sieberer}, and here we focus mostly on the differences: 1) We introduce a method to make the continuum limit mathematically rigorous; 2) We show that in order to preserve the normalization condition, one should retain the $t=t^{\prime}$ contribution in the bare Green's function.  

The Hamiltonian of the trivial one-site model reads $H=\mu c^{\dagger}c$,
where $\mu$ is the on-site energy and can be regarded as the chemical potential. The projective quantum jump operator is the particle number operator $c^\dagger c$. Thus, the Lindblad master equation describing the evolution under the Hamiltonian and the unconditional continuous projective measurements can be expressed as 
\begin{equation}
    \partial_{t}\rho=-i\left[H,\rho\right]+\gamma\left(c^{\dagger}c\rho c^{\dagger}c-\frac{1}{2}\left\{ c^{\dagger}c,\rho\right\} \right).
\end{equation}
This equation can be formally expressed as $\rho_{t_{f}}=\lim_{N\rightarrow\infty}\left(1+\delta_{t}\cdot\mathcal{L}\right)^{N}\rho_{0}$, where we have divide the time interval into $N$ slices, and  $\mathcal{L}$ is Liouvillian superoperator, which is defined as 
\begin{equation}
  \mathcal{L}\left(\rho\right)=-i\left[H,\rho\right]+\gamma\left(c^{\dagger}c\rho c^{\dagger}c-\frac{1}{2}\left\{ c^{\dagger}c,\rho\left(t\right)\right\} \right).
\end{equation}
Based on the recursion equation $\rho_{n+1}=\left(1+\delta_{t}\cdot\mathcal{L}\right)\rho_{n}$, one can get the final state $\rho_{t_f}$. In order to get the path integral based on the fermionic coherent state, we should first expand the density matrix in the fermionic coherent basis. Thus, we have
\begin{widetext}
\begin{equation}
    \rho_n=\int d\bar{\psi}_{+,n}d\psi_{+,n}d\bar{\psi}_{-,n}d\psi_{-,n}e^{-\bar{\psi}_{+,n}\psi_{+,n}}e^{-\bar{\psi}_{-,n}\psi_{-,n}}\langle\psi_{+,n}|\rho_{n}|-\psi_{-,n}\rangle|\psi_{+,n}\rangle\langle-\psi_{-,n}|,
\end{equation}
where $|\psi\rangle$ is the fermionic coherent state, and $\psi$, $\bar{\psi}$ are independent Grassmann numbers. We also have 
\begin{equation}
\begin{split}
    &\quad\langle\psi_{+,n+1}|\rho_{n+1}|\psi_{-,n+1}\rangle\\
    &=\int d\bar{\psi}_{+,n}d\psi_{+,n}d\bar{\psi}_{-,n}d\psi_{-,n}e^{\left(\bar{\psi}_{+,n+1}-\bar{\psi}_{+,n}\right)\psi_{+,n}}e^{\bar{\psi}_{-,n}\left(\psi_{-,n+1}-\psi_{-,n}\right)}\langle\psi_{+,n}|\rho_{n}|-\psi_{-,n}\rangle\\
    &\quad+\delta_{t}\int d\bar{\psi}_{+,n}d\psi_{+,n}d\bar{\psi}_{-,n}d\psi_{-,n}e^{-\bar{\psi}_{+,n}\psi_{+,n}}e^{-\bar{\psi}_{-,n}\psi_{-,n}}\langle\psi_{+,n+1}|\mathcal{L}\left(|\psi_{+,n}\rangle\langle-\psi_{-,n}|\right)|-\psi_{-,n+1}\rangle\langle\psi_{+,n}|\rho_{n}|-\psi_{-,n}\rangle.
\end{split}
\end{equation}
Since the Keldysh Lindblad partition function is defined as $Z=\tr(\rho_{t_f})$, we take the trace of $\rho_{n+1}$ in the fermionic coherent basis, and we have 
\begin{equation}
\begin{split}
    &\quad\tr(\rho_{n+1})\\
    &=\int\prod_{j=n}^{n+1}d\bar{\psi}_{+,j}d\psi_{+,j}d\bar{\psi}_{-,j}d\psi_{-,j}e^{\bar{\psi}_{-,n+1}\psi_{+,n+1}}e^{-\bar{\psi}_{+,n+1}\psi_{+,n+1}}e^{-\bar{\psi}_{-,n+1}\psi_{-,n+1}}e^{\left(\bar{\psi}_{+,n+1}-\bar{\psi}_{+,n}\right)\psi_{+,n}}e^{\bar{\psi}_{-,n}\left(\psi_{-,n+1}-\psi_{-,n}\right)}\\
    &\qquad\times\langle\psi_{+,n}|\rho_{n}|-\psi_{-,n}\rangle\\
    &\quad+\int\prod_{j=n}^{n+1}d\bar{\psi}_{+,j}d\psi_{+,j}d\bar{\psi}_{-,j}d\psi_{-,j}e^{\bar{\psi}_{-,n+1}\psi_{+,n+1}}e^{-\bar{\psi}_{+,n+1}\psi_{+,n+1}}e^{-\bar{\psi}_{-,n+1}\psi_{-,n+1}}e^{\left(\bar{\psi}_{+,n+1}-\bar{\psi}_{+,n}\right)\psi_{+,n}}e^{\bar{\psi}_{-,n}\left(\psi_{-,n+1}-\psi_{-,n}\right)}\\
    &\qquad\times\delta_{t}\left\{ -i\left[H\left(\bar{\psi}_{+,n+1},\psi_{+,n}\right)-H\left(\bar{\psi}_{-,n},\psi_{-,n-1}\right)\right]+\gamma\bar{\psi}_{+,n+1}\psi_{+,n}\bar{\psi}_{-,n}\psi_{-,n+1}{\color{red}-\frac{1}{2}}\gamma\left(\bar{\psi}_{+,n+1}\psi_{+,n}+\bar{\psi}_{-,n}\psi_{-,n+1}\right)\right\}\\
    &\qquad\times\langle\psi_{+,n}|\rho_{n}|-\psi_{-,n}\rangle.
\end{split}
\end{equation}
In Ref.~\cite{sieberer}, in the continuum limit, $\bar{\psi}_{+,n+1}\psi_{+,n}\bar{\psi}_{-,n}\psi_{-,n+1}$, $\bar{\psi}_{+,n+1}\psi_{+,n}$, and $\bar{\psi}_{-,n}\psi_{-,n+1}$ are directly set to $\bar{\psi}_{+}(t)\psi_{+}(t)\bar{\psi}_{-}(t)\psi_{-}(t)$, $\bar{\psi}_{+}(t)\psi_{+}(t)$, and $\bar{\psi}_{-,n}\psi_{-,n+1}$, respectively. Here, in order to make the continuum limit rigorous, we make those Grassmann numbers of the dissipation part be at the time argument following the procedure:
\begin{equation}\label{eq:MCLR}
\begin{split}
    &\quad\int d\bar{\psi}_{+,n}d\psi_{+,n}d\bar{\psi}_{-,n}d\psi_{-,n}e^{\bar{\psi}_{-,n}\left(\psi_{-,n+1}-\psi_{-,n}\right)}\bar{\psi}_{-,n}\psi_{-,n+1}\\
    &=\int d\bar{\psi}_{+,n}d\psi_{+,n}d\bar{\psi}_{-,n}d\psi_{-,n}e^{\bar{\psi}_{-,n}\left(\psi_{-,n+1}-\psi_{-,n}\right)}\bar{\psi}_{-,n}\left(\psi_{-,n+1}-\psi_{-,n}+\psi_{-,n}\right)\\
    &=\int d\bar{\psi}_{+,n}d\psi_{+,n}d\psi_{-,n} d\bar{\psi}_{-,n}\left[\frac{\delta}{\delta\bar{\psi}_{-,n}}e^{\bar{\psi}_{-,n}\left(\psi_{-,n+1}-\psi_{-,n}\right)}\right]\bar{\psi}_{-,n}+\int d\bar{\psi}_{+,n}d\psi_{+,n}d\bar{\psi}_{-,n}d\psi_{-,n}e^{\bar{\psi}_{-,n}\left(\psi_{-,n+1}-\psi_{-,n}\right)}\bar{\psi}_{-,n}\psi_{-,n}\\
    &=\int d\bar{\psi}_{+,n}d\psi_{+,n}d\bar{\psi}_{-,n}d\psi_{-,n}e^{\bar{\psi}_{-,n}\left(\psi_{-,n+1}-\psi_{-,n}\right)}\left(\bar{\psi}_{-,n}\psi_{-,n}+1\right).
\end{split}
\end{equation}
$\bar{\psi}_{+,n+1}\psi_{+,n}\bar{\psi}_{-,n}\psi_{-,n+1}$ and  $\bar{\psi}_{+,n+1}\psi_{+,n}$ can be treated in the same way.

Therefore, we have 
\begin{equation}\label{eq:KLPFiDV}
\begin{split}
    Z&=\frac{1}{\tr\left(\rho_{0}\right)}\int\prod_{j=0}^{1}d\bar{\psi}_{+,j}d\psi_{+,j}d\bar{\psi}_{-,j}d\psi_{-,j}\\
    &\times\exp\left\{ \begin{bmatrix}\bar{\psi}_{+,0} & \bar{\psi}_{+,1} & \bar{\psi}_{-,1} & \bar{\psi}_{-,0}\end{bmatrix}\begin{bmatrix}-1 & 0 & 0 & -\rho\\
h_{-} & -1 & 0 & 0\\
0 & 1 & -1 & 0\\
0 & 0 & h_{+} & -1
\end{bmatrix}\begin{bmatrix}\psi_{+,0}\\
\psi_{+,1}\\
\psi_{-,1}\\
\psi_{-,0}
\end{bmatrix}+ \gamma\delta_{t}\left[\bar{\psi}_{+,0}\psi_{+,0}\bar{\psi}_{-,0}\psi_{-,0}{\color{red}+\frac{1}{2}}\left(\bar{\psi}_{+,0}\psi_{+,0}+\bar{\psi}_{-,0}\psi_{-,0}\right)\right]\right\},
\end{split}
\end{equation}
where we choose $N=1$ for simplicity, $h_{\mp}=1\mp i\mu\delta_{t}$, $\rho=\langle\psi_{+,0}|\rho_{0}|-\psi_{-,0}\rangle$, and the initial state is chosen to be an exponential form, such as a thermal state. One finds that after doing the treatment shown in Eq.~\eqref{eq:MCLR}, the sign before the fator $1/2$ in the dissipation term is changed (see Eq.~\eqref{eq:MCLR} and Eq.~\eqref{eq:KLPFiDV}). Note that the dissipation part (the second term of the second line in Eq.~\eqref{eq:KLPFiDV}) depends on the same time argument, thus one can  directly take the continuum limit and this procedure is mathematically rigorous now. The Keldysh-Lakin-Ovchinnikov (KLO) transformation~\cite{kamenev} leads Eq.~\eqref{eq:KLPFiDV} to 
\begin{equation}
\label{eq:KLPF}
    Z=\frac{1}{\tr\left(\rho_{0}\right)}\int\prod_{j=0}^{1}d\bar{\psi}_{1,j}d\psi_{1,j}d\bar{\psi}_{2,j}d\psi_{2,j}\exp\left\{-\bar{\Psi}\left(-i\hat{G}^{-1}\right)\Psi+\gamma\delta_t\left[-\bar{\psi}_{1,0}\psi_{1,0}\bar{\psi}_{2,0}\psi_{2,0}+\frac{1}{2}\left(\bar{\psi}_{1,0}\psi_{2,0}+\bar{\psi}_{2,0}\psi_{1,0}\right)\right]\right\},
\end{equation}
where $\bar{\Psi}=\begin{bmatrix}\bar{\psi}_{1,0} & \bar{\psi}_{1,0} & \bar{\psi}_{2,0} & \bar{\psi}_{2,0}\end{bmatrix}$, $\Psi=\begin{bmatrix}\psi_{1,0} & \psi_{1,0} & \psi_{2,0} & \psi_{2,0}\end{bmatrix}^t$, $\psi_{a,j}$ is the Grassmann number after the KLO transformation with $a\in\{1,2\}$ being the Keldysh indices and $j\in\{0,1\}$ being the discrete time indices, and 
\end{widetext}
\begin{equation}
    -i\hat{G}^{-1}=-\frac{1}{2}\begin{bmatrix}-\rho & -h_{+} & h_{+} & -2+\rho\\
h_{-} & -1 & -3 & h_{-}\\
h_{-} & -1 & 1 & h_{-}\\
-2-\rho & h_{+} & -h_{+} & \rho
\end{bmatrix}.
\end{equation}
The bare Green's function in the discrete time version then reads
\begin{equation}
\begin{split}
    i\hat{G}&=\left(-i\hat{G}^{-1}\right)^{-1}\\
    &=\begin{bmatrix}\frac{1}{2} & 0 & \frac{1-\rho}{1+\rho}h_{+} & \frac{1-\rho}{1+\rho}\\
h_{-} & \frac{1}{2} & \frac{1-\rho}{1+\rho} & \frac{1-\rho}{1+\rho}h_{-}\\
0 & 0 & -\frac{1}{2} & 0\\
0 & 0 & -h_{+} & -\frac{1}{2}
\end{bmatrix}+\frac{1}{2}\begin{bmatrix}0 & 0 & 0 & 1\\
0 & 0 & 1 & 0\\
0 & 1 & 0 & 0\\
1 & 0 & 0 & 0
\end{bmatrix}.
\end{split}
\end{equation}
In the continuum limit, Eq.~\eqref{eq:KLPF} can be expressed as 
\begin{equation}\label{eq:BFGF}
    i\hat{G}\left(t,t^{\prime}\right)=\begin{bmatrix}iG_0^{R}\left(t,t^{\prime}\right) & iG_0^{K}\left(t,t^{\prime}\right)\\
0 & iG_0^{A}\left(t,t^{\prime}\right)
\end{bmatrix}+\frac{1}{2}\begin{bmatrix}0 & 1\\
1 & 0
\end{bmatrix}\delta_{t,t^{\prime}},
\end{equation}
where $\delta_{t,t^{\prime}}$ should be interpreted as the Kronecker symbol.
In the standard Keldysh field theory~\cite{kamenev}, people usually omit the term proportional to $\delta_{t,t^{\prime}}$ in Eq.~\eqref{eq:BFGF}, and only keep the first term of Eq.~\eqref{eq:BFGF}.
In order to check the normalization condition $Z=1$, one can expand Eq.~\eqref{eq:KLPF} in powers of $\gamma$, and treat each order with the help of Wick's theorem. 
In our problem here, one will find that this  $\delta_{t,t^{\prime}}$ term has to be kept so as to preserve the normalization, as one will encounter the equal-time correlation: $\langle\psi_{2}(t)\bar{\psi}_1(t)\rangle$.

Generalizing to the model considered in the main text, one can obtain the Keldysh Lindblad partition function Eq.~(3) of the main text.\\

\section{Keldysh Treatment of Disordered Fermionic systems and Comparison with Our Problem}\label{sec:KToDFS}

The Keldysh treatment of the disordered fermionic system can be found in Ref.~\cite{horbach,kamenev}, and here we just quote some discussions connected with our problem.

In the traditional studying of the disordered fermionic system or the weak localization effect, one usually assume a static and spatial-dependent disorder potential $V_{dis}(\mathbf{x})$ through the disorder action 
\begin{equation}
    S_{dis}[V_{dis}]=\int dx\,V_{dis}(\mathbf{x})\,\bar{\psi}_a(x)\hat{\tau}_0^{ab}\psi_b(x),
\end{equation}
where the configuration of $V_{dis}(\mathbf{x})$ satisfies the Gaussian distribution and thus the disorder averaging takes the form
\begin{equation}
  \langle\cdots\rangle_{dis}=\int\mathcal{D}[V_{dis}]\exp\left\{-\pi\nu\tau_{el}\int d\mathbf{x}\,V_{dis}^2(\mathbf{x})\right\}\cdots,
\end{equation}
where $\tau_{el}$ is the elastic scattering time. 
Performing the disorder averaging for $\exp(iS_{dis})$, one can get 
\begin{widetext}
\begin{equation}
\begin{split}
    \langle e^{iS_{dis}}\rangle_{dis}&=\int\mathcal{D}[V_{dis}]\exp\left\{-\int d\mathbf{x}\,\pi\nu\tau_{el}V_{dis}^2(\mathbf{x})+iV_{dis}(\mathbf{x})\int dt\;\bar{\psi}_a(\mathbf{x},t)\hat{\tau}_0^{ab}\psi_b(\mathbf{x},t)\right\},\\
    &=\exp\left\{-\frac{1}{4\pi\nu\tau_{el}}\int d\mathbf{x}\int dtdt^{\prime}\;\bar{\psi}_a(\mathbf{x},t)\psi_a(\mathbf{x},t)\bar{\psi}_b(\mathbf{x},t^{\prime})\psi_b(\mathbf{x},t^{\prime})\right\}.
\end{split}
\end{equation}
And then, the partition function after disorder averaging reads
\begin{equation}\label{eq:PTfDFS}
    Z=\int\mathcal{D}[\psi]\exp\left\{iS_0-\frac{1}{4\pi\nu\tau_{el}}\int d\mathbf{x}\int dtdt^{\prime}\;\bar{\psi}_a(\mathbf{x},t)\psi_a(\mathbf{x},t)\bar{\psi}_b(\mathbf{x},t^{\prime})\psi_b(\mathbf{x},t^{\prime})\right\},
\end{equation}
where $S_0$ is the free-fermion action. Note that the disorder averaging introduce a four-fermion term into the action.

For convenience, we also put the Keldysh Lindblad partition function of our problem here:
\begin{equation}\label{eq:KLPFoMFFG}
    Z=\int\!\mathcal{D}\left[\psi\right]\exp\left\{iS_{0} -\frac{\gamma}{2}\int d\mathbf{x}dt \pmb{[}\bar{\psi}_{a}\left(\mathbf{x},t\right)\psi_{a}\left(\mathbf{x},t\right)\bar{\psi}_{b}\left(\mathbf{x},t\right)\psi_{b}\left(\mathbf{x},t\right)-\bar{\psi}_{a}\left(\mathbf{x},t\right)\hat{\tau}_{1}^{ab}\psi_{b}\left(\mathbf{x},t\right)
  \pmb{]}\right\}.
\end{equation}
\end{widetext}
Comparing these two equations,  Eq.~\eqref{eq:PTfDFS} and Eq.~\eqref{eq:KLPFoMFFG}, one can observe that the two four-fermion terms are in a similar form. Thus, in some sense, these two different problems are unified in the framework of the functional Keldysh field theory. Here, we emphasize again that such a similarity is not obvious in the master equation formalism, and can {\em only} be found when one resorts to the Keldysh path integral formalism (see Fig. 2 in the main text).

\begin{figure}[t]
	\centerline{\includegraphics[height=6.5cm]{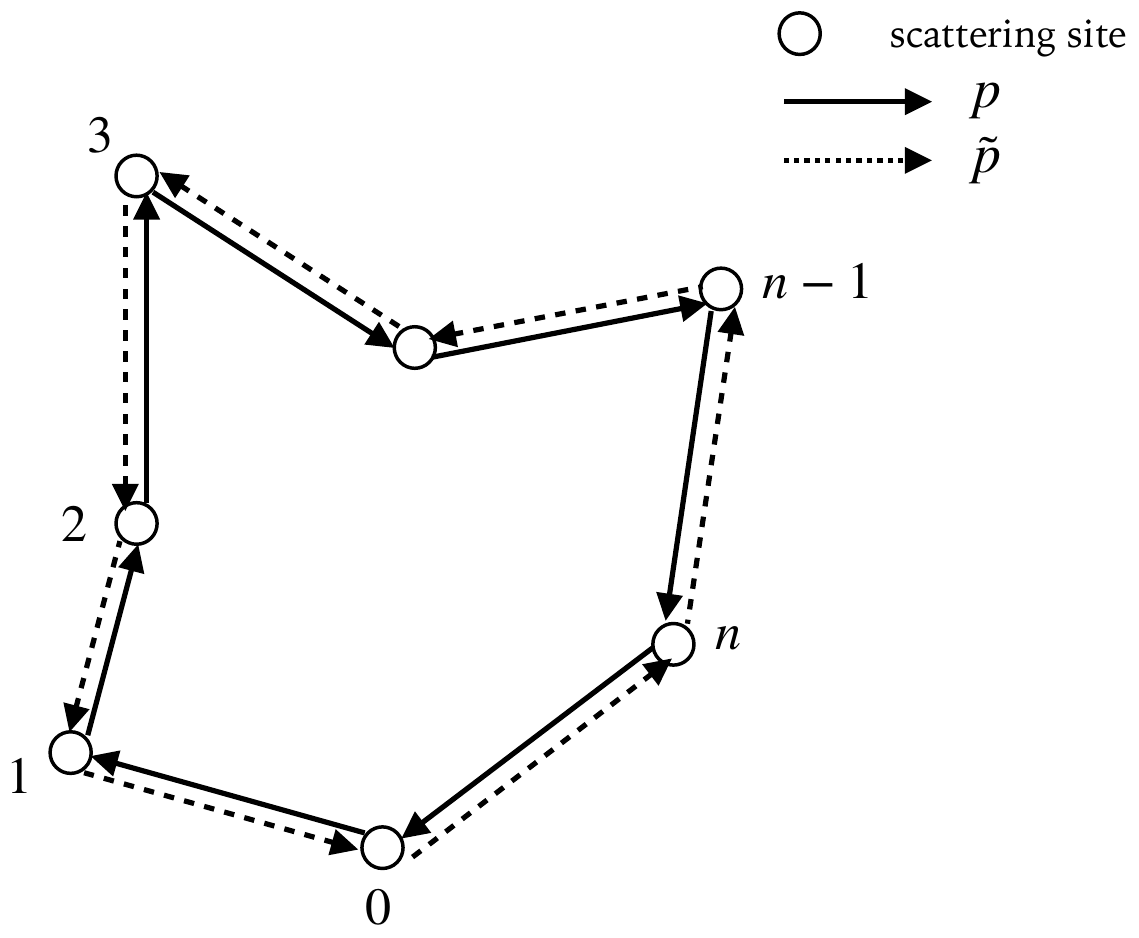}}
	\caption{Scattering of an electron along two time-reversed paths in the disordered fermionic systems. 	\label{fig:TI}}
\end{figure}

In the disordered fermionic systems, the time-reversal symmetry (TRS) is present, thus the remarkable weak localization exists. A rudimentary conceptual understanding of the weak localization is that it results from the {\em constructive interference} of two {\em time-reversed} paths of an electron~\cite{coleman}. Consider the amplitude of an electron to return to its starting point. In general, it will encounter a sequence of scattering sites (see Fig. \ref{fig:TI}), which are manifestations of disorders. For each path $p$, there is a time-reversed path $\tilde{p}$ when the system is time-reversal symmetric. Amplitudes of the electron around $p$ and $\tilde{p}$ are the same, while for other paths, phases should be random. Therefore, the electron will have a quantum mechanically enhanced probability of returning to its starting point due to the time-reversal symmetry. From this argument, we know that the time-reversal symmetry is significant to the weak localization in the disordered case. Theoretically, the time-reversal symmetry will result in another set of soft modes, known as {\em Cooperons}~\cite{horbach,altland,kamenev}, in the effective theory---Keldysh nonlinear sigma model. Those Cooperon modes will provide an infrared divergent correction to the DC conductivity, and then imply a localization transition from the metallic phase. If the system do not have time-reversal symmetry, the constructive interference of two time-reversed paths does not exist, and thus the weak localization disappears. Theoretically, the Cooperon modes disappear due to the lack of TRS, and only diffuson modes contribute to the transport properties. As a consequence, the DC conductivity will present in the Drude form without divergent corrections. 

In the measurement case or the dephasing system, the time-reversal symmetry is obviously absent. Thus the soft modes analogous to Cooperons in the disordered systems do not exist in the measurement case or the dephasing case. By this comparison, we know that diffusons are the only dominant excitations or soft modes in the Keldysh nonlinear sigma model, and there will not be localization for weak measurement strength ($\gamma\ll \epsilon_F$). In Sec. \ref{sec:NV} of the main text, we provide a numerical example to support our theory and verify our predictions. From the numerics (see Fig. \ref{fig:TI}), we find that for all finite measurement strengths, the weak localization does not exist. And for large $\gamma$, the scaling of the diffusion coefficient (and thus the conductivity) is also $1/\gamma$. The scaling behavior implies that for an {\em infinite} measurement strength, the system will be localized. However, this is a manifestation of the quantum Zeno effect instead of the weak localization. As the measurement strength $\gamma$ can be regarded as the number of measurement invents in a unit time interval ($\gamma\sim 1/\tau$ with $\tau$ being the duration of one-shot measurement), $\gamma\rightarrow\infty$ means that the system is measured all the time. In this sense, $\gamma\rightarrow\infty$ corresponds to the quantum Zeno limit.  \\

\section{Bosonic Effective Theory and the Keldysh Nonlinear Sigma Model}\label{sec:BEFaKNSM}

\subsection{Bosonic Effective Theory}
Following the procedure introduced in Sec.~\ref{sec:LMEtKFT}, one can get the Keldsyh Lindblad partition function for our problem, that is Eq. \eqref{eq:KMPF} in the main text.
For the four-fermion term in the dissipation part, we introduce an auxiliary {\em time-local} bosonic field $\hat{Q}$ to decouple it with the help of the identity
\begin{equation}
\begin{split}\label{eq:IfQ}   \hat{1}&=\int\mathcal{D}\left[\hat{Q}\right]\exp\left[-\frac{\gamma}{2}\left(\pi\nu\right)^{2}\tr\left(\hat{Q}^{2}\right)\right]\\ 
    &=\int\mathcal{D}\left[\hat{Q}\right]\exp\left[-\frac{\gamma}{2}\left(\pi\nu\right)^{2}\int dx\;\hat{Q}^{ab}\left(x\right)\hat{Q}^{ba}\left(x\right)\right].
\end{split}
\end{equation}

The definition of $\hat{Q}$ is similar with the definition of an operator in quantum mechanics. In quantum mechanics, an operator $\hat{\mathcal{O}}$ in the position basis can be expressed as $\hat{\mathcal{O}}=\int d\mathbf{x_1}d\mathbf{x_2}\hat{\mathcal{O}}(\mathbf{x}_1,\mathbf{x}_2)|\mathbf{x}_1\rangle\langle\mathbf{x}_2|$, and $\hat{\mathcal{O}}(\mathbf{x}_1,\mathbf{x}_2)$ is the matrix element of $\hat{\mathcal{O}}$. If $\hat{\mathcal{O}}$ is diagonal in the position basis, then $\hat{\mathcal{O}}$ reduces to $\hat{\mathcal{O}}=\int d\mathbf{x}\hat{\mathcal{O}}(\mathbf{x})|\mathbf{x}\rangle\langle\mathbf{x}|$. More often, the matrix element $\hat{\mathcal{O}}(\mathbf{x}_1,\mathbf{x}_2)$ is just a number. However, one can always generalize it to the case that the matrix element is also a matrix. Such a generalization is widely used in quantum field theories and tensor network methods. In our work, the definition of $\hat{Q}$ is exactly such a generalization. The non-zero matrix element of $\hat{Q}$ which is diagonal in the position basis, can be expressed as 
\begin{equation}
    \hat{Q}(x)=\left[\begin{array}{ll}Q^{11}(x) & Q^{12}(x) \\ Q^{21}(x) & Q^{22}(x)\end{array}\right].
\end{equation}
After using this definition, the trace over $\hat{Q}$ becomes the trace over both the Keldysh space and the spacetime basis. For example, $\tr\left(\hat{Q}^2\right)$ in Eq. \eqref{eq:IfQ} is defined as 
\begin{widetext}
\begin{equation} 
\begin{split}
\tr\left(\hat{Q}^2\right)&=\tr_K\left[\int dx \langle x|\int dx_1dx_2\hat{Q}(x_1)|x_1\rangle\langle x_1|\cdot\hat{Q}(x_2)|x_2\rangle\langle x_2|x\rangle\right]\\
&=\tr_K\left[\int dx \langle x|\int dx_1\hat{Q}^2(x_1)|x_1\rangle\langle x_1|x\rangle\right]\\
&=\tr_K\left[\int dx\hat{Q}^2(x)\right]\\
&=\int dx \hat{Q}^{ab}(x)\hat{Q}^{ba}(x),
\end{split}
\end{equation}
where $\tr_K$ stands for the trace over the Keldysh space, $a,b\in\{1,2\}$ are Keldysh indices and repeated indices imply summation. 

After introducing the auxiliary fields $\hat{Q}$, one arrives at 
\begin{equation}
\begin{split}   Z&=\int\mathcal{D}\left[\hat{Q}\right]\mathcal{D}\left[\psi\right]\exp\left\{iS_{0} -\frac{\gamma}{2}\left(\pi\nu\right)^{2}\tr\left(\hat{Q}^{2}\right)-\gamma\int dx\left[\pi\nu\bar{\psi}_{a}\left(x\right)\hat{Q}^{ab}\left(x\right)\psi_{b}\left(x\right)-\frac{1}{2}\bar{\psi}_{a}\left(x\right)\hat{\tau}_{1}^{ab}\psi_{b}\left(x\right)\right]\right\}.
\end{split}
\end{equation}
Using the Gaussian integration, one arrives at the effective bosonic theory depending only on $\hat{Q}$:
\begin{equation}
    Z=\int\mathcal{D}\left[\hat{Q}^{\prime}\right]\exp\left\{ -\frac{\gamma}{2}\left(\pi\nu\right)^{2}\tr\left[\left(\hat{Q}^{\prime}+\frac{1}{2\pi\nu}\hat{\tau}_{1}\right)^{2}\right]+\tr\ln\left[-i\hat{G}^{-1}+\gamma\pi\nu\hat{Q}^{\prime}\right]\right\},
\end{equation}
where we have let $\hat{Q}^{\prime}=\hat{Q}-\frac{1}{2\pi\nu}\hat{\tau}_1$, and $\hat{Q}^{\prime}$ is still Hermitian. In the following, we will relabel $\hat{Q^{\prime}}$ as $\hat{Q}$ again.  For higher orders ($\ge 2$) of the expansion in powers of $\gamma$, we can replace $\hat{G}$ with $\hat{G}_0$ (the first term in Eq.~\eqref{eq:BFGF}) due to the fact that the $t=t^{\prime}$ line is only a manifold of measure zero~\cite{kamenev}. Then, one gets the Keldysh Lindblad Partition function shown in Eq.~(5) of the main text:
\begin{equation}\label{eq:EBT1}
    Z=\int\mathcal{D}\left[\hat{Q}\right]\exp
    \left\{ -\frac{\gamma}{2}\left(\pi\nu\right)^{2}\tr\left[\hat{Q}^{2}+\left(\frac{1}{2\pi\nu}\hat{\tau}_{1}\right)^{2}\right]+\tr\ln\left[-i\hat{G}^{-1}_{0}+\gamma\pi\nu\hat{Q}\right]\right\}.
\end{equation}
%Note that for $\gamma\ll\epsilon_F$ (this condition is also used to deriving the saddle point), this result is also valid up to the first order of $\gamma$. 
\end{widetext}
\subsection{Time-Local Keldysh Nonlinear Sigma Model}

Taking the variation over $\hat{Q}(x)$, one gets the saddle point equation of the action in Eq.~\eqref{eq:EBT1}, and one can check that the constant configuration $\hat{\Lambda}=\frac{1}{2\pi\nu}\hat{\tau}_3$ satisfies the saddle point equation. For large-scale physics, we just focus on the massless fluctuation, which can be generated by $\hat{Q}(x)=\hat{\mathcal{R}}^{-1}(x)\hat{\Lambda}\hat{\mathcal{R}}(x)$. Note that now $\hat{Q}(x)$ is constrained by $\hat{Q}^2(x)=(\frac{1}{2\pi\nu})^2\hat{\tau}_0$. And then one finds that only the $\tr\ln$ term in Eq.~\eqref{eq:EBT1} will contribute to the dynamics, while other terms only contribute some constants. Thus, in the following, we can just focus on the $\tr\ln$ term.

Note that the bare Green's function $\hat{G}_0$ can be expressed as $\hat{G}_0=\hat{\mathcal{U}}^{-1}\hat{G}_{0d}\,\hat{\mathcal{U}}$, where 
\begin{equation}
\begin{split}  &\hat{\mathcal{U}}^{-1}= \hat{\mathcal{U}}=\sum_{\epsilon}\begin{bmatrix}1 & F_{\epsilon}\\
0 & -1
\end{bmatrix}|\epsilon\rangle\langle\epsilon|,\\
&\hat{G}_{0d}=\sum_{\mathbf{k},\epsilon}\begin{bmatrix}G_0^R(\mathbf{k},\epsilon) & 0\\
0 & G_0^A(\mathbf{k},\epsilon)
\end{bmatrix}|\mathbf{k},\epsilon\rangle\langle\mathbf{k},\epsilon|,
\end{split}
\end{equation}
and $F_{\epsilon}=1-2n_F(\epsilon)$ with $n_F(
\epsilon)$ being the Fermi-Dirac distribution function. Thus, the statistical information is actually encoded in the matrix $\hat{\mathcal{U}}$. The statistical distribution in $\hat{G}_0$ comes from the initial thermal state $\rho_0=\exp[-\beta\sum_{
\mathbf{k}}c_{\mathbf{k}}^{\dagger}(\epsilon_{\mathbf{k}}-\epsilon_F)c_{\mathbf{k}}]$. We would like to obtain an effective theory depending only on $\hat{Q}$ to describe the  physics of our problem. To this end, we first make a similarity transformation to encode the statistical information in $\hat{Q}$ instead. Note that due to the cyclic property of the trace operation, this similarity does not change the theory. And then the $\tr\ln$ term in Eq.~\eqref{eq:EBT1} now becomes $\tr\ln\left[-i\hat{G}^{-1}_{0d}+(\gamma/2)\hat{Q}\right]$, where $\hat{Q}$ is redefined as $\hat{Q}=\hat{\mathcal{U}}^{-1}\hat{\mathcal{R}}^{-1}\hat{\tau}_3\hat{\mathcal{R}}\hat{\mathcal{U}}$, and in the spacetime basis, $\hat{G}_{0d}^{-1}=i\partial_t+\frac{\nabla^2}{2m}+\epsilon_F+i0\hat{\tau}_3$. Therefore, we have
\begin{widetext}
\begin{equation}
\begin{split}
    iS\left[\hat{Q}\right]&=\tr\ln\left[-i\hat{G}_{0d}^{-1}+(\gamma/2)\hat{\mathcal{U}}^{-1}\hat{\mathcal{R}}^{-1}\hat{\Lambda}\hat{\mathcal{R}}\hat{\mathcal{U}}\right]\\
    &=\tr\ln\left\{\left[-i\hat{\mathcal{R}}\left(i\partial_{t}+\frac{\nabla^{2}}{2m}+\epsilon_{F}\right)\hat{\mathcal{R}}^{-1}-i\hat{\mathcal{U}}^{-1}i0\hat{\tau}_{3}\hat{\mathcal{U}}\right]+\frac{\gamma}{2}\hat{\tau}_{3}\right\}\\
    &\approx\tr\ln\left[\hat{\mathcal{G}}^{-1}+i\hat{\mathcal{U}}^{-1}\hat{\mathcal{R}}\left(\partial_{t}\hat{\mathcal{R}}^{-1}\right)\hat{\mathcal{U}}+i\hat{\mathcal{U}}^{-1}\hat{\mathcal{R}}\left(\mathbf{v}_{F}\cdot\mathbf{\nabla}\hat{\mathcal{R}}^{-1}\right)\hat{\mathcal{U}}\right],
\end{split}
\end{equation}
\end{widetext}
where $\hat{\mathcal{G}}^{-1}=i\partial_{t}+\frac{\nabla^{2}}{2m}+\epsilon_{F}+i\frac{\gamma}{2}\hat{\mathcal{U}}\hat{\tau}_{3}\hat{\mathcal{U}}$, and $\mathbf{v}_F\cdot\nabla$ comes from the linearization of the dispersion relation near the Fermi energy: $\mathbf{k}^2/(2m)-\epsilon_F\approx\mathbf{v}_F\cdot\mathbf{k}\rightarrow-i\mathbf{v}_F\cdot\mathbf{k}$~\cite{horbach,kamenev}. Note that the saddle point configuration $\propto\hat{\tau}_3$ plays the role of the self-energy. In the energy-momentum basis, we have 
\begin{equation}
    \hat{\mathcal{G}}(\mathbf{k},\epsilon)=\hat{\mathcal{U}}_{\epsilon}\begin{bmatrix}\frac{1}{\epsilon-\xi_{\mathbf{k}}+i\gamma/2} & 0\\
0 & \frac{1}{\epsilon-\xi_{\mathbf{k}}-i\gamma/2}
\end{bmatrix}\hat{\mathcal{U}}_{\epsilon},
\end{equation}
where $\xi_{\mathbf{k}}=\mathbf{k}^2/(2m)-\epsilon_F$. Expanding the $\tr\ln$ term in powers of $\partial_t\hat{\mathcal{R}}^{-1}$ and $\nabla\hat{\mathcal{R}}^{-1}$ (similar with the Taylor expansion of the function $\ln(1+x)$), one will arrive at the time-local Keldysh nonlinear sigma model
\begin{equation}
iS\left[\hat{Q}\right]=\pi\nu\tr\left[\partial_{t}\hat{Q}\right]-\frac{1}{4}\pi\nu D\tr\left[\left(\nabla\hat{Q}\right)^{2}\right],
\end{equation}
where $\partial_t\hat{Q}\equiv\partial_t(\hat{\mathcal{U}}\hat{\mathcal{R}}^{-1})\hat{\tau}_3\hat{\mathcal{R}}\hat{\mathcal{U}}$. The linear order of the spatial gradient is zero due to the angular integration. A similar calculation can be found in Chapter 11 of Ref.~\cite{kamenev}. As our auxiliary field $\hat{Q}$ only depends on one time argument, while in the disordered fermionic case, the auxiliary field depends on two time variables, here we term our Keldysh nonlinear sigma model as the time-local KNSM to distinguish from the KNSM in the disordered fermionic case.
The Keldysh nonlinear sigma model in the presence of the vector potential can be derived from those similar calculations.

\end{appendix}

%%%%%%%%%%%%%%%%%%%%
%%%%%%%%%%%%%%%%%%%%
\bibliography{KNSM_Ref}

\end{document}